\documentclass[12pt]{article}

\usepackage{graphicx}
\usepackage{scicite}
\usepackage{times}

\newcommand{\citep}[1]{\cite{#1}}
\newcommand{\citet}[1]{\cite{#1}}

\newcommand{\vp}[3]{\makebox[.6\width][r]{
\raisebox{-1.5ex}{$#1^{+#2}_{-#3}$}} } 
\renewcommand{\vp}[3]{$#1^{+#2}_{-#3}$}

\newcommand{\sgl}[1]{\makebox[.6\width][r]{
\raisebox{-1.5ex}{$#1$}}}

\newenvironment{sciabstract}{%
\begin{quote} \bf}
{\end{quote}}

\newpage

\newcounter{lastnote}

\title{Cosmological Constraints from Strong Gravitational Lensing in Clusters of Galaxies}

\author
{Eric Jullo,$^{1,2}$ Priyamvada Natarajan,$^{3\ast,4}$ Jean-Paul Kneib,$^{2}$ \\
Anson D'Aloisio,$^{4}$ Marceau Limousin,$^{2,5}$ Johan Richard,$^{6}$ and Carlo Schimd$^{2}$\\
\\
\normalsize{$^{1}$Jet Propulsion Laboratory, California Institute of Technology, Pasadena, CA 91109}\\
\normalsize{$^{2}$Laboratoire d'Astrophysique de Marseille, CNRS, Universite de Provence, 
38 rue Frederic Joliot-Curie,} \\ 
\normalsize{13388 Marseille Cedex 13, France}\\
\normalsize{$^{3}$Department of Astronomy, Yale University, P. O. Box 208101, 
New Haven, CT 06511-8101, USA}\\
\normalsize{$^{4}$Department of Physics, Yale University, P. O. Box 208120, 
New Haven, CT 06520-8120, USA}\\
\normalsize{$^{5}$Dark Cosmology Centre, Niels Bohr Institute, University of Copenhagen,}\\ 
\normalsize{Juliane Maries Vej 30, 2100 Copenhagen, Denmark}\\
\normalsize{$^{6}$Institute for Computational Cosmology, Department of
Physics, Durham University,} \\ 
\normalsize{South Road, Durham, DH1 3LE, UK}\\ \\
\normalsize{$^\ast$To whom correspondence should be addressed; E-mail:priyamvada.natarajan@yale.edu.}
}

\date{}

\begin{document}

\maketitle


\baselineskip24pt

\begin{sciabstract}
{\bf Current efforts in observational cosmology are focused on
characterizing the mass-energy content of the Universe. We present
results from a geometric test based on strong lensing in galaxy
clusters. Based on {\it Hubble Space Telescope} images and extensive
ground-based spectroscopic follow-up of the massive galaxy cluster
Abell~1689, we used a parametric model to simultaneously constrain the
cluster mass distribution and dark energy equation of state. Combining
our cosmological constraints with those from X-ray clusters and the
{\it Wilkinson Microwave Anisotropy Probe} 5-year data gives
$\Omega_{\rm m} = 0.25 \pm 0.05$ and $w_{\rm x} = -0.97 \pm 0.07$
which are consistent with results from other methods. Inclusion of our
method with all other techniques available brings down the current
$2\sigma$ contours on the dark energy equation of state parameter
$w_{\rm x}$ by about 30\%.}
\end{sciabstract}

Measurements of the Hubble diagram for Type Ia supernovae
\citep{riess1998,perlmutter1998,tonry2003, riess2004} combined with
constraints from the Wilkinson Microwave Anisotropy Probe ({\it
WMAP5}) \citet{spergel2007,hinshaw2008}, cosmic shear observations
\citet{bacon2000,kaiser2000,vanwaerbeke2000,wittman2000,
semboloni2006}, cluster baryon fractions \citet{allen2004}, cluster
abundances \citet{vikhlinin09} and baryon acoustic oscillations (BAO)
from galaxy surveys \citet{efstat2002,seljak2005,eisenstein2005}
suggests that $\sim72$\% of the total energy density of the Universe
is in the form of an unknown constituent with negative pressure - the
so-called dark energy, that powers the measured accelerating
expansion.  These observations probe the equation-of-state parameter
$w_{\rm x}$, defined as the ratio of pressure to energy density,
through its effect on the expansion history and structure of the
Universe. The current goal of cosmology is to understand the
properties of dark energy by placing tighter constraints on its
equation of state. In the currently favored flat $\Lambda$CDM
model \footnote{The currently favored `concordance' cosmological model
that best describes the Universe is the $\Lambda$CDM paradigm in which
the bulk of the matter and energy density are dominated by dark matter
and dark energy with baryons contributing only $\sim$ 5\%.}, dark
energy is attributed to a cosmological constant, for which $w_{\rm x}
= -1$. Type Ia supernovae, baryon acoustic oscillations, cluster
abundances and cosmic shear appear to be very promising techniques to
tighten constraints on the equation-of-state parameter in the near
future. Because all of these techniques have biases, systematics and
degeneracies, it is only in combination that robust estimates of
cosmological parameters can be obtained.

The most recent census from a combination of techniques suggests that
$\sim 72\%$ of the energy density in the Universe is in the form of
dark energy that is powering the accelerating expansion of the
Universe. In the progression towards
In this work, we present results from a technique that exploits the
strong gravitational lensing of distant background galaxies by massive
galaxy clusters. Through their effect on the local space-time
geometry, massive foreground structures cause the deflection and
shearing of light rays originating from distant sources. In the case
of strong lensing, the light beams are deflected so strongly that they
can often result in the observation of several distorted images of a
given single background galaxy. The positions of these multiple images
depend strongly on the detailed properties of the lens mass
distribution \citet{kneib1996,natarajan1998,smith2005}. Because the
image positions also depend on the angular diameter distance ratios
between the lens, source and observer, they encapsulate information
about the underlying cosmology. We capitalize on this dependence on
the geometry to derive constraints on the cosmological parameters
$\Omega_{\rm m}$ (the mean matter density) and $w_{\rm x}$.

Constraining the energy content of the Universe using multiple sets of
arcs in cluster lenses has been explored in the past
\citet{paczynski1981,link1998,cooray1999,golse2002a,sereno2002,soucail2004,
gilmore2009}. In particular, simultaneous inversion of the lens and derivation
of cosmological constraints can be performed based on the cosmological sensitivity 
of the angular size-redshift relation with sources at distinct redshifts
\citet{link1998}. In this method, the angular
diameter distance ratios for 2 images from different sources defines
the `family ratio' $\Xi$, from the cosmological dependence of which
constraints on $\Omega_{\rm m}$ and $w_{\rm x}$ are extracted:
\begin{equation}
    \Xi(z_{\rm l},z_{\rm s1},z_{\rm s2};\Omega_{\rm M},\Omega_{\rm X},w_{\rm X})=\frac{D(z_{\rm l},z_{\rm
s1})}{D(0,z_{\rm s1})}\frac{D(0,z_{\rm s2})}{D(z_{\rm l},z_{\rm
s2})}
\end{equation}
where $z_{\rm l}$ is the lens redshift, $z_{\rm s1}$ and $z_{\rm
s2}$ are the two source redshifts, and $D(z_{1},z_{2})$ is the
angular diameter distance.

Application of this method to the cluster Abell 2218 using 4 multiple image systems at
distinct redshifts, resulted in $\Omega_{\rm m}<0.37$ and
$w_{\rm x} <-0.80$ for a flat Universe \citet{soucail2004}. A
recent feasibility study demonstrates that the
degeneracies of this technique are entirely distinct from those of other
cluster methods and that combining the results from several simulated
clusters with $> 10$ multiple image families in each can provide a
powerful probe of dark energy \citet{gilmore2009}. 

We have applied this technique to the massive, lensing cluster
Abell~1689 at redshift $z=0.184$.  Based on images from the Advanced
Camera for Surveys ({\it ACS}) aboard the Hubble Space Telescope ({\it
HST}) this cluster has 114 multiple images from 34 unique background
galaxies, 24 of which have secure spectroscopic redshifts (ranging
from $z \sim 1$ to $z \sim 5$) obtained with the Very Large Telescope
(VLT) and Keck Telescope spectrographs
\citet{broadhurst2005,limousin2007b}. Abell~1689 is amongst the
richest clusters in terms of the number density of galaxies in its
core. It is also amongst the most luminous of galaxy clusters in X-ray
wavelengths, with an absolute X-ray luminosity of $L_X = 20.74 \times\
10^{37}\ {\rm W}$ \citet{ebeling1996}. Observationally, Abell~1689
consists of two groups of galaxies~: a dominant one located at the
center coincident with the peak of X-ray emission, and a secondary
concentration about 1 arcminute North-East of the main one. Studies
have shown that this second northern group is at a slightly higher
redshift, suggesting thus that these two groups might actually be
merging \citet{czoske2004}. The projected mass enhancement produced by
such a merging configuration, could therefore explain the stunningly
large number of multiple images identified in this cluster. Previous
work shows that the mass distribution of Abell~1689 is well modeled
with a set of parameterized elliptical pseudo-isothermal lensing
potentials \citet{limousin2007b}. We utilized the most recent
parametric model of Abell~1689, which is able to reproduce the
observed image configurations to within an average positional accuracy
of $2.87$ arcseconds, assuming a $\Lambda$CDM cosmology. We
solved the lens equation in the source plane for Abell~1689 as it is
computationally efficient.  Inverting in the lens (image) plane
provides additional information but is computationally prohibitive at
present \citet{limousin2007b}. Our simplified model that has a total
of 21 free parameters consists of two large-scale potentials, a
galaxy-scale potential for the central brightest cluster galaxy (BCG),
and includes the modeling of $58$ of the brightest cluster
galaxies. Therefore, we explicitly include the effect of substructure
in the lens plane and assigned potentials associated with bright
cluster galaxies. The velocity dispersion and scale radii of all but
one (the BCG) of the cluster galaxies were assumed to follow
empirically motivated scaling relations, which have been previously
utilized to model cluster lenses \citet{natarajan1997,natarajan2009}.

Despite the large number of multiple images observed, not all of them
can be utilized to constrain cosmology. From the initial 114 images,
we only used those (i) with robust, measured spectroscopic redshifts;
and (ii) excluded those in the regions of the cluster with low $S/N$
in the mass reconstruction.  This selection results in the culling of
multiple images that lie in the most uncertain regions of the mass
distribution. Moreover, we identified several bright spots in some
well resolved multiple images, which we used to increase the number of
families. Applying criterion (i), resulted in a catalog of 102 images.
Imposing criteria (i) and (ii), we obtained a catalog of 28 images
arising from 12 families all with measured spectroscopic redshifts
(Fig.~\ref{fig:a1689_clines}), providing a total of 32 constraints. We assume flatness as a prior
$(\Omega_{\rm tot} = 1)$ and fix the Hubble parameter at $H_0 = 74\,
\mathrm{km/s/Mpc}$ \citet{suyu2010}, as our cosmography test
is not sensitive to the value of the Hubble parameter.

For each of the observed image systems with $n$ images, we determine
the goodness of fit for a particular set of model parameters using a
source plane $\chi^2$,
\begin{equation}
\chi^2 = \sum_{i=1}^{n}{\frac{ \left[M \left(\vec{\beta}_i - \left< \vec{\beta} \right>\right) \right]^2}{\sigma^2_i}},
\end{equation}
where $\vec{\beta}_i$ is the source plane position corresponding to
image $i$, $\left< \vec{\beta} \right>$ is the family barycenter, M is
the magnification tensor, and $\sigma_i$ is the total (observational
and modeling) error. The total $\chi^2$ was obtained by summing over
families and was used in conjunction with a Markov Chain Monte Carlo
(MCMC) sampler to probe the posterior probability density function
(PDF) as a function of all relevant model parameters \citet{jullo2007}
(SOM). The key degeneracies with cosmological parameters for this
technique arise from the velocity dispersions, ellipticity and core
radii of the large scale mass clumps in the model (Fig.~S4).

The angular resolution of {\it HST} images is on the order of $0.1$
arcseconds. However, the modeling errors are generally larger due to
the complexity of the cluster mass distribution, as well as the effect
of intervening structures along the line of sight. We quantified the
errors due to the presence of structure along the line of sight using
the Millennium Simulation halo catalogs \citet{millenium}. We
quantified the errors on an image by image basis. By randomly slicing
through snapshots of the simulation and tiling them at redshift planes
between the observer and the source, we constructed $1000$
line-of-sight realizations. We then ray traced through each
realization with the Abell~1689 model included to estimate the effect
of intervening halos on image positions. In most cases, these
line-of-sight halos perturbed image positions but did not alter the
multiplicity of the images. These perturbations induce positional
displacements of the order of 1 arcsecond.  Therefore, about 1
arcsecond of the error between observed image positions and model
image predictions can be attributed to the presence of structure along
the line of sight behind the cluster. The presence of projected
correlated and associated large-scale structure (filaments) increases
the cross-section to strong gravitational lensing making these
clusters more efficient lenses, however, simulations show that this is
a sub-dominant effect to that of unassociated distant large-scale
structure. Therefore, the presence of aligned correlated large scale
structure at the redshift of the cluster does not scupper the recovery
of cosmological parameters \footnote{Filaments aligned at finite
inclinations behind the cluster (thus breaking azimuthal symmetry) do
yield larger deviations between the projected and multi-plane models
compared to the symmetric case. For the non-symmetric case, the
deviations are typically on the same level as the errors due to
scatter in the cluster galaxy properties. These deviations are still
subdominant to the effects of uncorrelated line-of-sight (LOS)
halos.}.

Earlier work on reconstructing the mass distribution of several
massive lensing clusters, shows that the association of dark matter
substructures with the locations of the brightest cluster galaxies is
well matched by that derived from the Millenium Simulation
\citet{natarajan2007a,natarajan2009}.  This supports our use of
luminosity-mass scaling relations to map substructure in the cluster.

The second potential source of error arises from modeling
uncertainties for the substructure in the lens plane.  This is
likely due to scatter in the assumed scaling relations for the
velocity dispersions and scale radii of cluster galaxies. Although the
mean correlations between these variables may be well described by
simple scaling relations, individual galaxies can deviate
substantially from them, introducing errors into the parameter
recovery. In order to quantify these modeling errors, we performed Monte
Carlo simulations of the lens system assuming a 20\% scatter in the
galaxy scaling relations.  This scatter induced modeling
errors on the background galaxy image positions that in some cases
were as large as $\sim1$ arcsecond. Therefore, the estimated errors from
substructure effects in the lens plane and along the line of sight are
comparable for the selected multiple image families in Abell~1689.

We use the catalog of 102 images (including images with photometric
redshift estimates) and our estimates of the observational and
modeling errors to obtain the marginalised PDF in the $\Omega_{\rm
m}-w_{\rm x}$ plane. Adding in quadrature the systematic errors
identified above to the positional uncertainties, results in a best
model for Abell~1689 with a well defined though broad degeneracy
between $\Omega_{\rm m}$ and $w_{\rm x}$. The ''concordance'' model of
$\Omega_{\rm m} \sim 0.3$ and $w_{\rm x} \sim -1$ lies within the
1$\sigma$ contour, but the way the degeneracy is pushed against the
prior limits suggests either a bias in the mass modeling, or
misidentified images (Fig.~S5). Our simulations show that the errors
in photometric redshift determination methods at present are too large
and limit the efficiency of our technique and introduce biases (see
SOM). Therefore, we excluded all images with photometric redshifts in
our modeling, bringing down the number of image families used from 34
to 24. Even with this cut, some images were badly reproduced with very
high RMS positional deviations, in particular, those that lie in
complex crowded regions of the cluster (regions where the
signal-to-noise of the mass map is low due to the presence of several
bright cluster galaxies in close proximity). These outlier images from
16 families highlight complex regions in the lens mass distribution,
not handled adequately by our simple parametric model. The outlier
image systems also tend to have higher redshifts; thus they are likely
to be more affected by uncertainties in modeling the intervening line
of sight structure as well.  We thus deliberately disregarded images
that lie in the most complex regions in the mass distribution.  While
the recovery of cosmological parameters is insensitive to the choice
of profile, both the PIEMD (Pseudo-Isothermal Elliptical Mass Distribution) 
and NFW (Navarro-Frenk-White) provide
comparable constraints on $w_{\rm x}$ and $\Omega_{\rm m}$ (Fig~S3),
observationally some clusters are better fit to one or the other
model. Abell 1689 is best fit with a PIEMD profile. The final culled
image catalog thus contains 28 images, derived from 12 families at
redshifts ranging from $z_S = 1.15$ to $z_S = 4.86$, all of which are
spectroscopically measured (Table~S1 and Fig.~\ref{fig:a1689_clines}).

Optimising our model with all the spectroscopically selected images,
including outlier images does not result in significant constraints on
either $\Omega_{\rm m}$ and $w_{\rm x}$ and yields an averaged reduced
$\chi^2 = 0.08$. This indicates that the line of sight and scaling
relation errors are likely overestimated. Thus, we optimized again but
this time excluding the outlier images. In this iteration, we obtained
an averaged reduced $\chi^2 \sim 28$, indicating that now the errors
were somewhat underestimated (Fig.~\ref{fig:combALL}). Owing to the
large estimates for the modeling errors, constraints obtained in this
fashion from a single cluster lens are fairly modest. However,
combining our results with those from X-ray clusters and {\it WMAP5}
leads to $\Omega_{\rm m}\,=\,0.25 \pm 0.05$ and $w_{\rm x}\,=\,-0.97
\pm 0.07$ (Fig.~\ref{fig:combALL}), which is consistent with the
values derived from combining {\it WMAP5} with SN and BAO,
$\Omega_{\rm m}\,=\,0.265 \pm 0.16 \pm 0.025$ and $w_{\rm x}\,=\,-0.96
\pm 0.06 \pm 0.12$ \citet{kessler2009}.

Our results when combined with the results from {\it WMAP5}
\citet{komatsu2009}, the supernovae ``Gold sample'' \citet{riess2004},
SNLS project \citet{astier2006} and SNEssence, and the BAO peak from
{\it SDSS} \citet{eisenstein2005}, give $0.23 < \Omega_{\rm m} < 0.33$
and $-1.12 < w_{\rm x} < -0.82$ at the 99\% confidence level
(Fig.~\ref{fig:clusALL}). This combination of all current viable
probes brings down the overall error in $w_{\rm x}$ by about
30\%. Therefore, the combination of cluster methods with {\it WMAP5}
has comparable potency to the combination of other cosmological
probes.

%%%%%%%%%%%%%% REFERENCES %%%%%%%%%%%%%%%%%%%

\bibliographystyle{Science}
\bibliography{all}

\noindent{40. EJ and JPK acknowledge fruitful discussions with D.~Coe,
L.~Koopmans, L.~Moustakas and M.~Auger. The authors are thankful for
computational resources from IDRIS/CNRS, JPL and J.-C Bourret. EJ
acknowledges a NASA post-doctoral fellowship; and PN acknowledges a
Radcliffe Fellowship at Harvard. JPK, ML and CS acknowledge support
from CNRS; and ML thanks CNES \& Ville de Marseille. The authors thank
Alexey Vikhlinin for providing data on cosmological constraints from
X-ray clusters to combine with our strong lensing results.}

\newpage

\begin{figure}[h]
\includegraphics[width=0.7\textwidth]{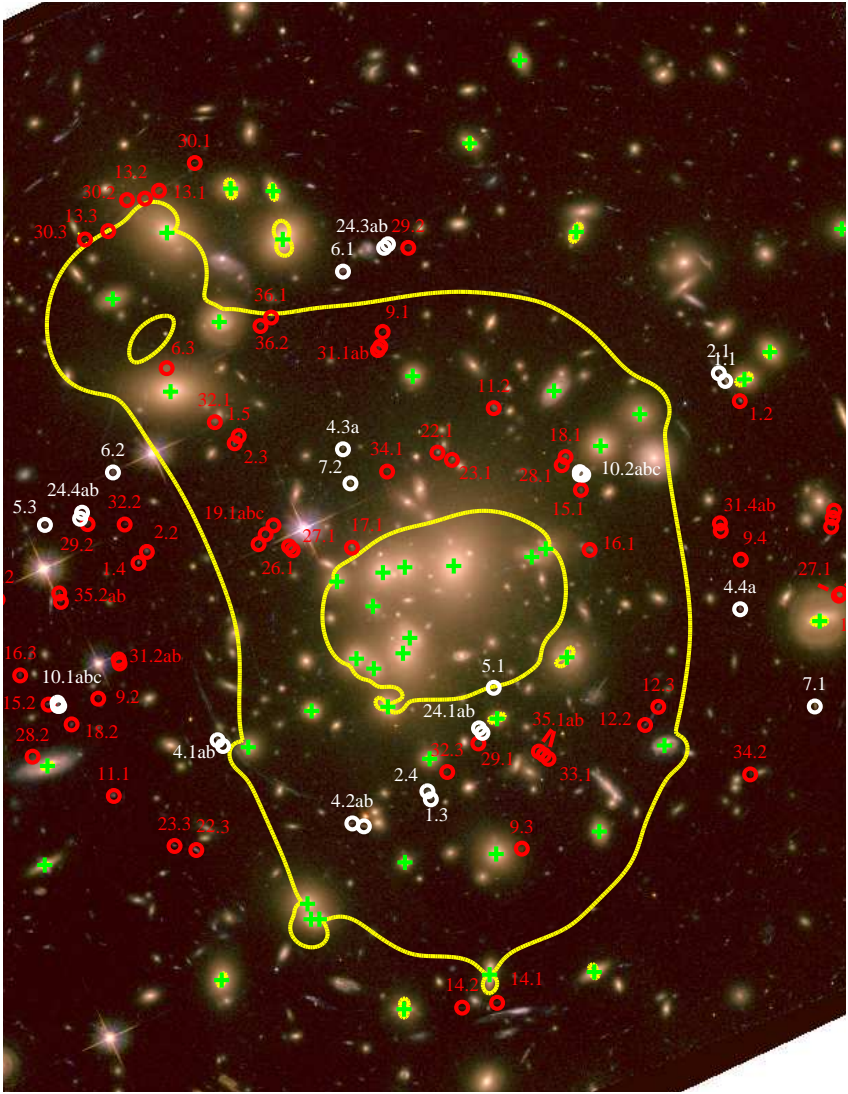} 
\caption{\label{fig:critline} The critical lines for a source at $z =
3$ are overplotted in yellow on the {\it HST ACS} image of
Abell~1689. The lensing mass model used is the one from which
we derived cosmological constraints. In addition to 2 large scale clumps and 
the BCG, this model includes the contribution of 58 cluster galaxies. The positions
of cluster galaxies are marked with green crosses. Overplotted in
white are the 28 multiple images arising from 12 families that we 
used in this work; the red circles mark the positions of the
rejected images.} 
\label{fig:a1689_clines} 
\end{figure}

\begin{figure}[h]
\begin{center}
\resizebox{7.0cm}{!}{\includegraphics{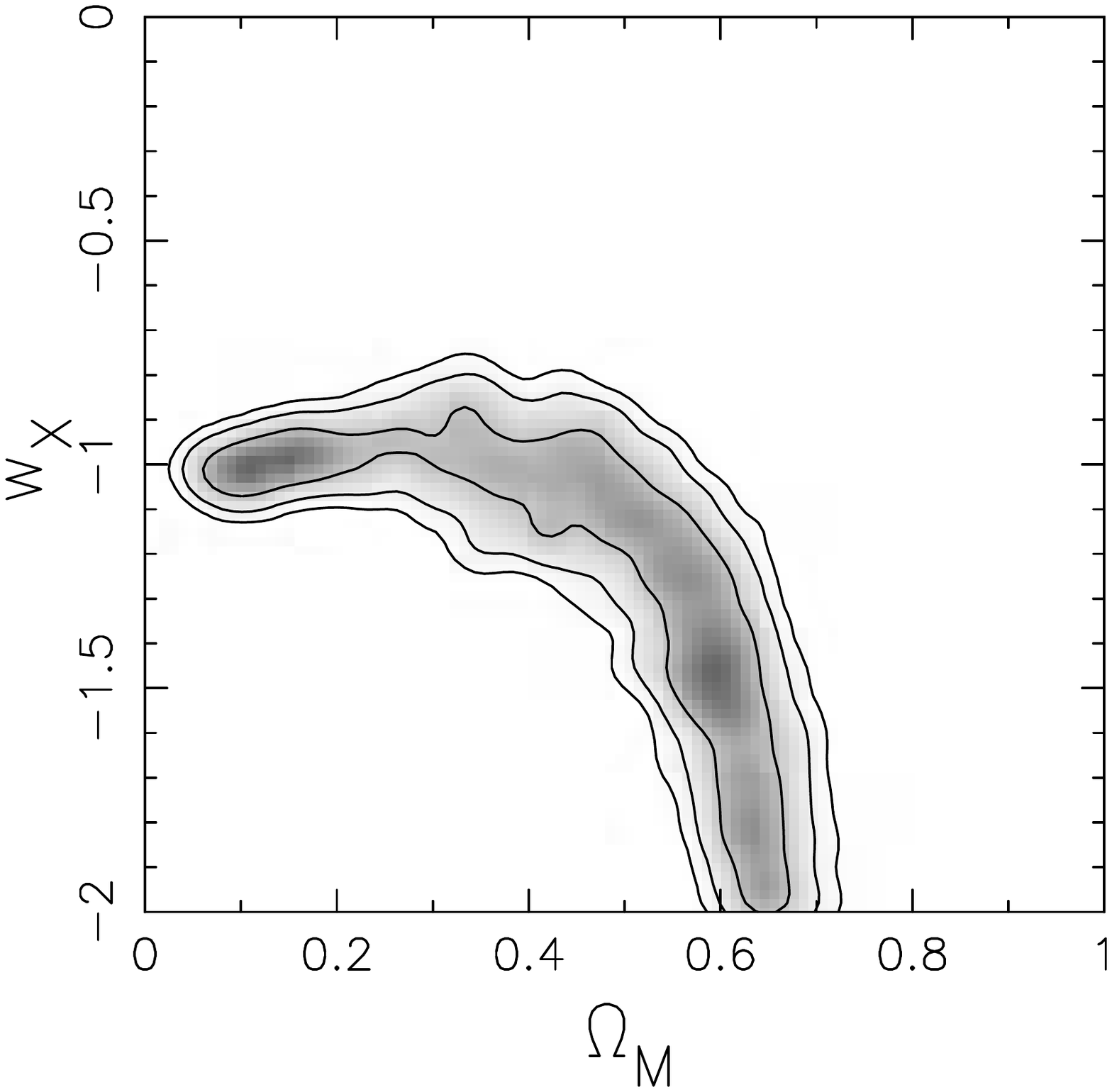}} \hspace{0.13cm} 
\resizebox{7.0cm}{!}{\includegraphics{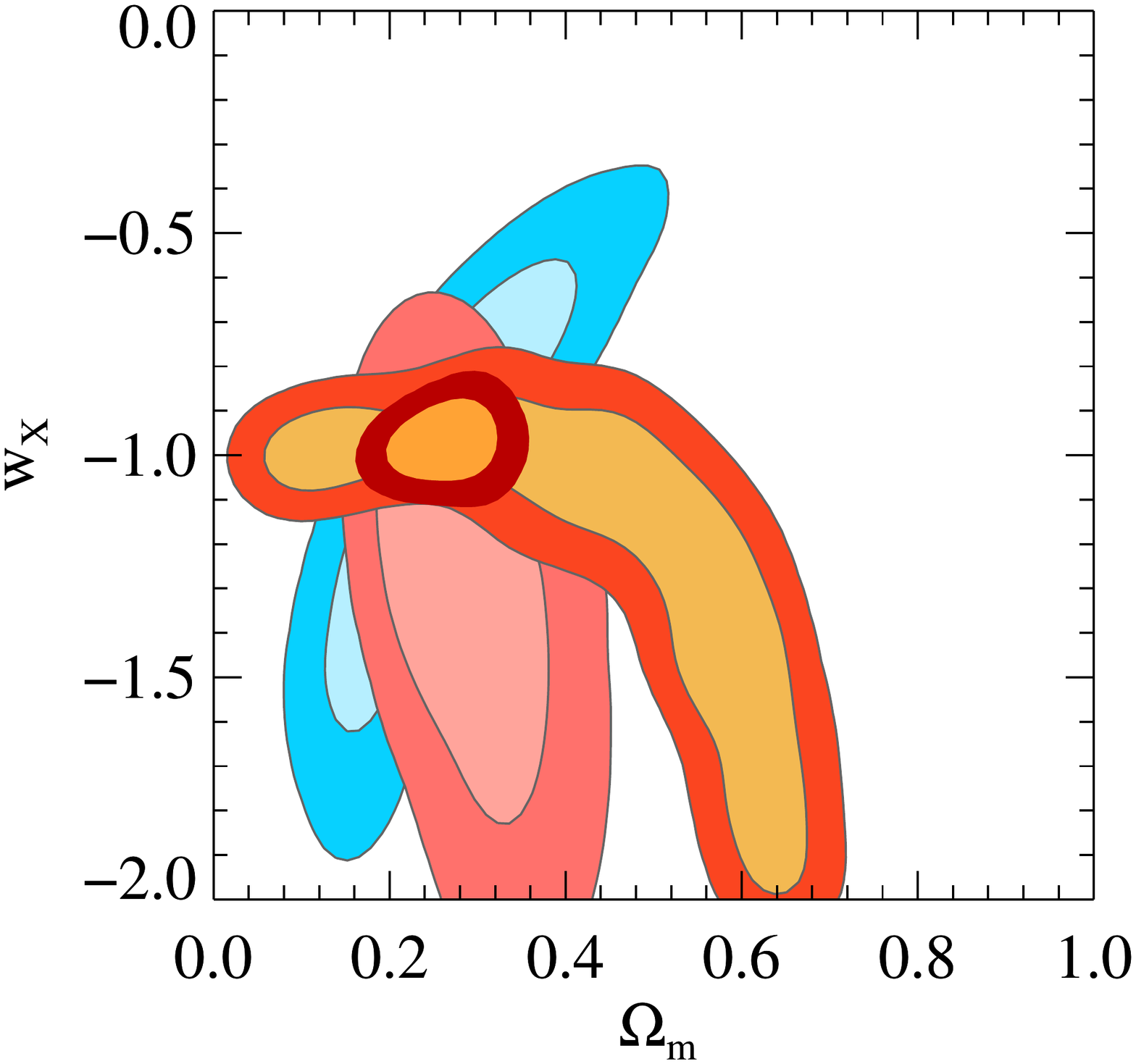}}
\end{center}
\caption{Left) The results from the simultaneous Bayesian
optimization of the detailed mass distribution and cosmological
parameters in the $\Omega_{\rm m}-w_{\rm x}$ plane for Abell~1689
using the 28 multiple images belonging to 12 families at distinct
redshifts as constraints from strong lensing including only
observational errors. The plotted contours are the 1, 2 and 3-$\sigma$ 
confidence levels. (Right) The results from combining
cosmological constraints from {\it WMAP5}+ evolution of X-ray clusters 
+cluster strong lensing (cluster only methods); the 1 and 2$\sigma$ contours
are plotted, blue contours - constraints from {\it WMAP5}, pink
contours - X-ray clusters, orange contours - cluster strong
lensing. We multiplied the likelihoods for the various techniques to 
obtain this plot. The degeneracy directions for X-ray
clusters and cluster strong lensing are orthogonal.}
\label{fig:combALL}
\end{figure}

\begin{figure}[h]
\begin{center}
\includegraphics[width=0.8\linewidth]{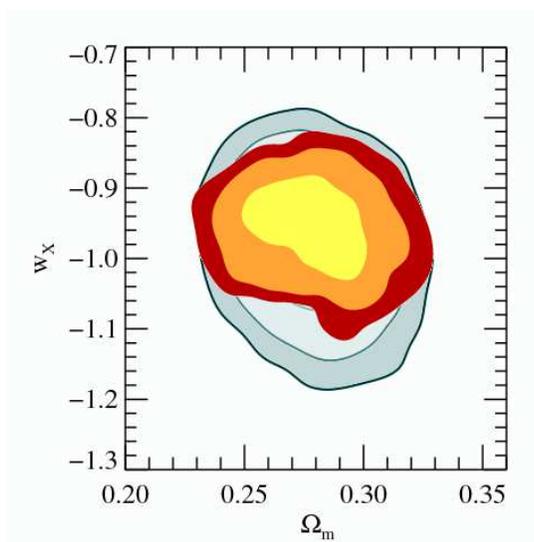}
\end{center}
\caption{Combination of constraints from
strong lensing, the {\it WMAP5} data \citet{komatsu2009}, the
supernovae ``Gold sample'' \citet{riess2004}, SNLS project
\citet{astier2006} and SNEssence, and the BAO peak from {\it SDSS}
\citet{eisenstein2005}. Contours for non-SL cosmological probes come
from the {\it WMAP} plotter. The overplotted contours are all 1, 2 and 
3-$\sigma$ confidence levels.}
\label{fig:clusALL}
\end{figure}

\clearpage

\vfill
\eject

\newpage

\topmargin 0.0cm
\oddsidemargin 0.2cm
\textwidth 16cm 
\textheight 21cm
\footskip 1.0cm

\section*{\bf Supporting Online Materials}

\section*{Mass reconstruction of Abell~1689}

As a starting point for the cosmography work, we use the strong
lensing based mass model presented in Limousin et~al. 2007. We
summarize this model briefly below. The mass distribution in
Abell~1689 was reconstructed in a parametric fashion using detected
strong lensing features from deep {\it HST ACS} observations and
extensive ground based spectroscopy. We find 36 multiply imaged
systems yielding 102 images, of which 24 systems have reliably
determined spectroscopic redshifts. These systems span a redshift
range between $z = 1.1$ and $z = 4.9$ and provide a total of 136
constraints for the analysis. The multiple images used and their
relevant properties, including location and redshift are shown in
Table~1. The resulting mass distribution in this cluster is found to
be bimodal: the dominant component is a central large scale mass
clump/halo coincident with the location of the brightest cluster
galaxy (BCG) and the center of the X-ray emission. The second mass
component is a another large scale mass clump/halo in the north-east
part of the {\it ACS} field, it also contributes substantially to the
overall mass distribution and is required by the data to reproduce
several of the multiple image geometries. In addition, this model
includes the galaxy scale dark matter haloes associated with the all
the identified cluster members (totalling 267) detected within the
{\it ACS} field providing the smaller scale structure in the dark
matter distribution. The critical lines for this mass model are shown
in Figure~\ref{fig:a1689_clines}. To keep the mass model simple and
yet include the contribution of small scale structure, standard
scaling laws with luminosity are assumed for the cluster
galaxies. Thus the final mass model has 33 free parameters. The large
number of observational constraints used (136 constraints) makes
Abell~1689 the most reliably reconstructed cluster to date, with a
mass distribution constrained at the per cent level, a level required
in order to exploit Abell~1689 efficiently as a cosmological probe.
{\bf Note that this is the baseline input mass model that we started with,
eventually, to extract cosmological parameters we pared down the model
to include only 58 of the cluster galaxies and only 28 images all with
measured spectroscopic redshifts.} Details of the selection are
presented in the paper.
\begin{figure}
\includegraphics[width=0.7\linewidth]{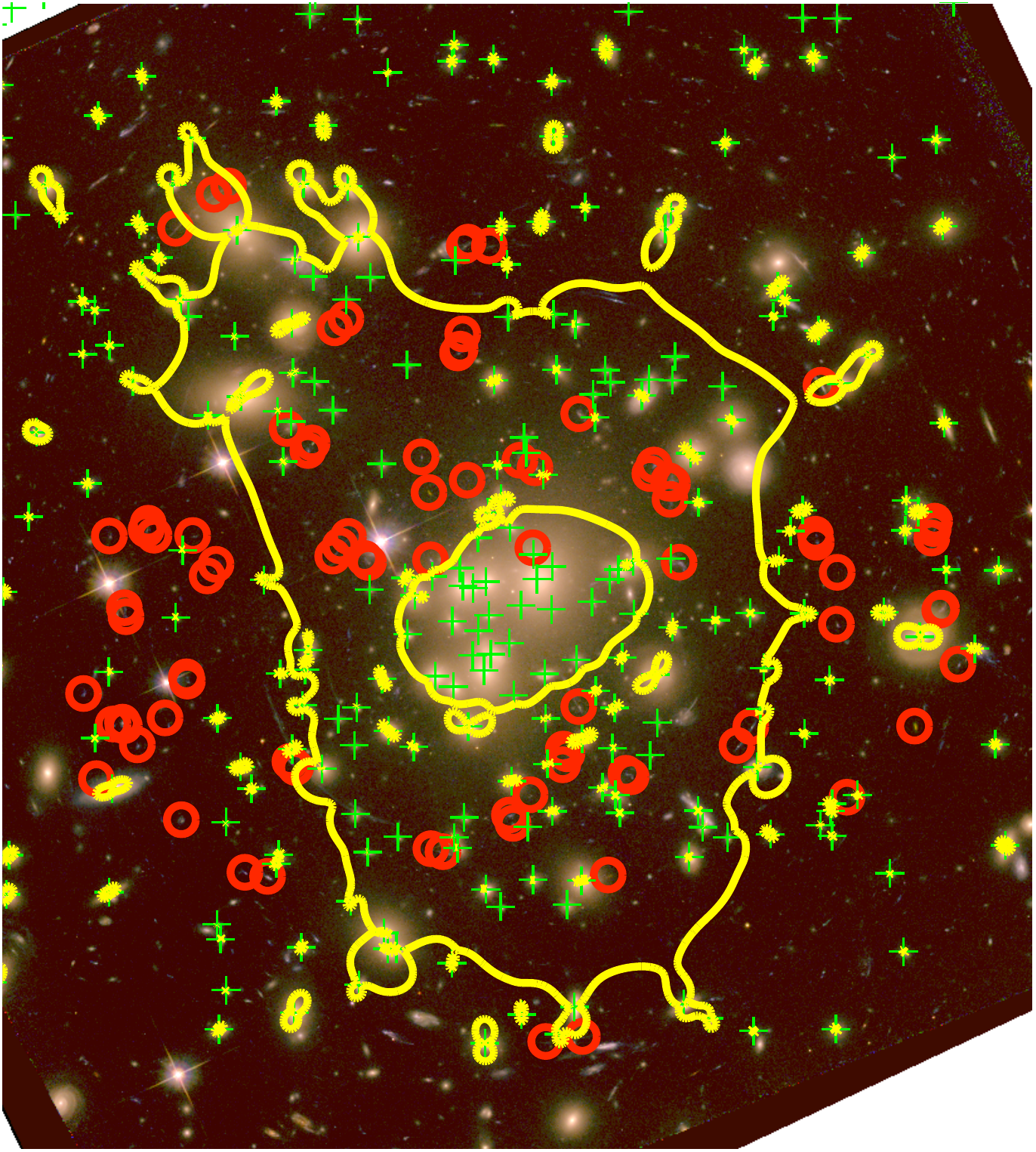}\\
\caption{Mass model for Abell~1689 published in Limousin et
al. (2007): the critical lines for a source at $z = 3$ are overplotted
in yellow on the {\it HST ACS} image of Abell~1689. This is done for
the best-fit lensing mass model that includes 267 cluster
galaxies. Overplotted in red are all the multiple images including
those with photo-metric redshifts. Cluster members selected in Abell~1689 
using the color-magnitude relation are marked with green crosses.}
\label{fig:a1689_clines}
\end{figure}

\begin{figure}
\centering
\begin{tabular}{c@{}c@{}c@{}c@{}c}
\includegraphics[width=0.20\linewidth]{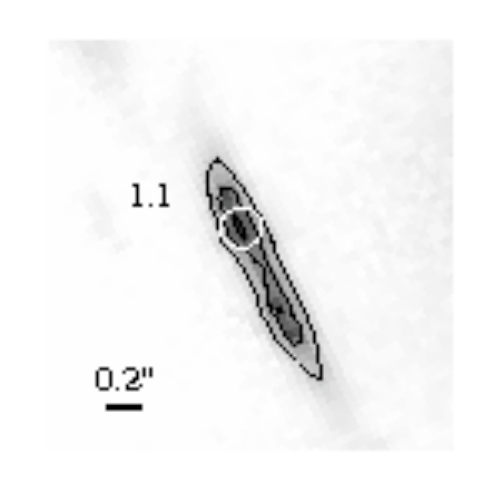}&
\includegraphics[width=0.20\linewidth]{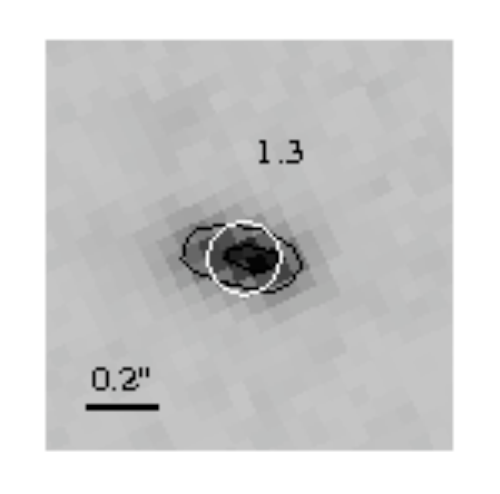}&
\includegraphics[width=0.20\linewidth]{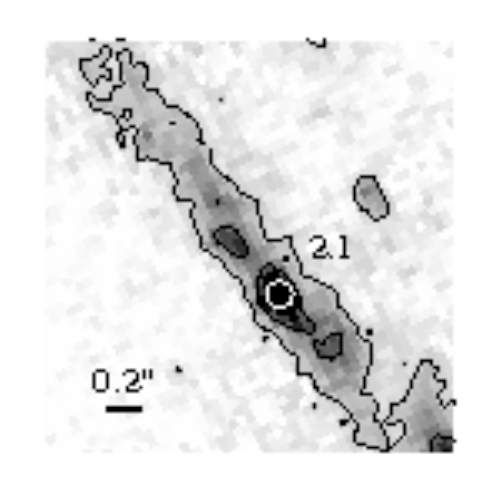}&
\includegraphics[width=0.20\linewidth]{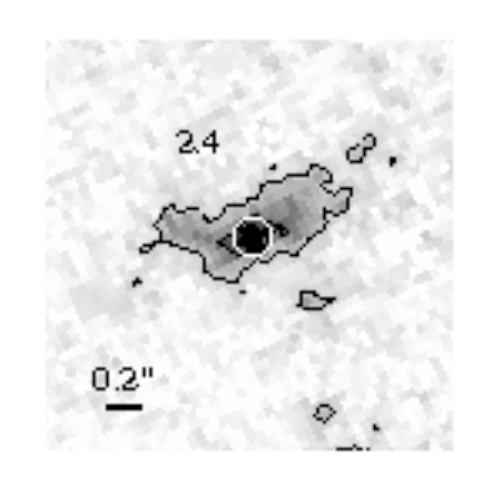}&
\includegraphics[width=0.20\linewidth]{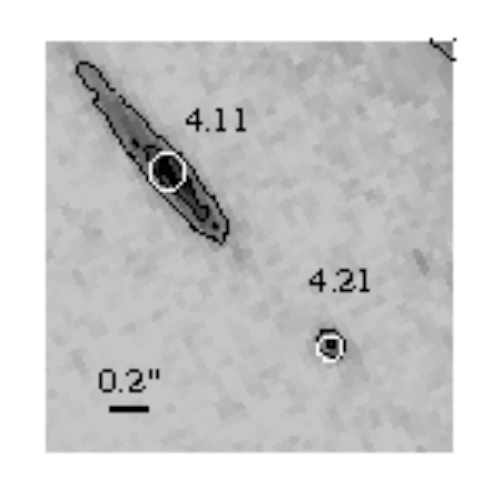}\\
\includegraphics[width=0.20\linewidth]{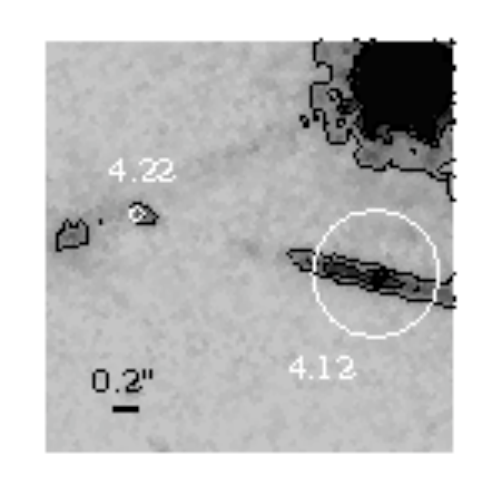}&
\includegraphics[width=0.20\linewidth]{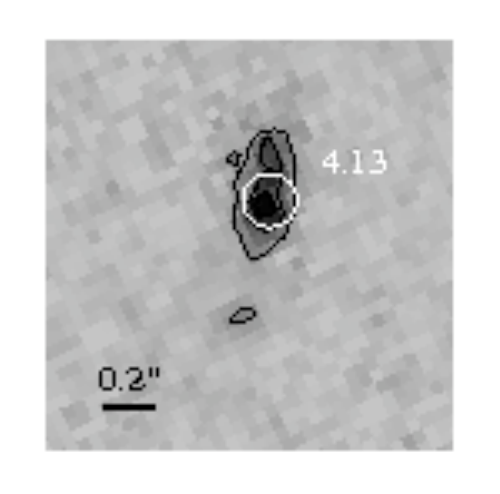}&
\includegraphics[width=0.20\linewidth]{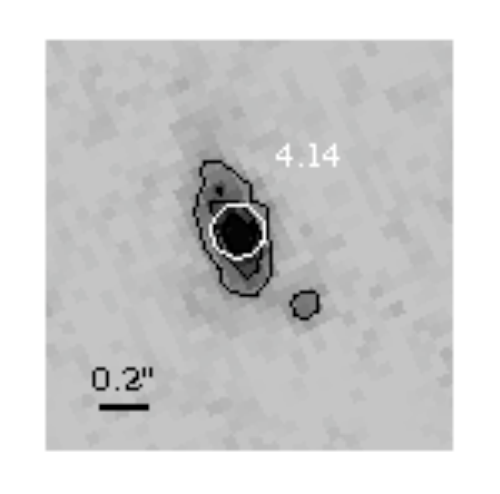}&
\includegraphics[width=0.20\linewidth]{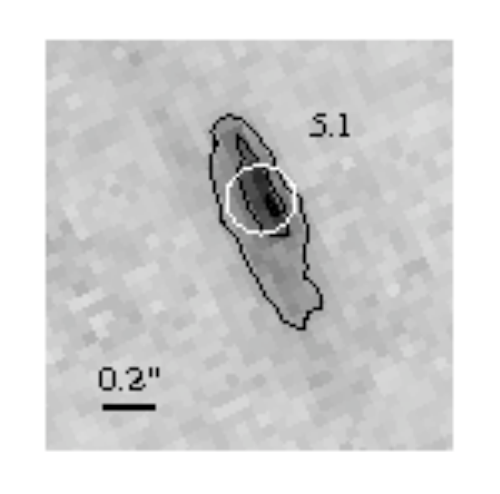}&
\includegraphics[width=0.20\linewidth]{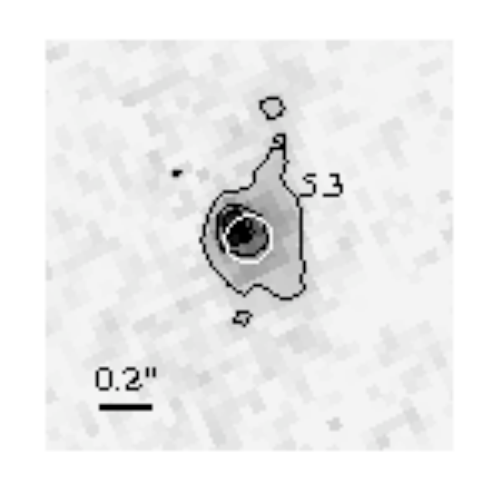}\\
\includegraphics[width=0.20\linewidth]{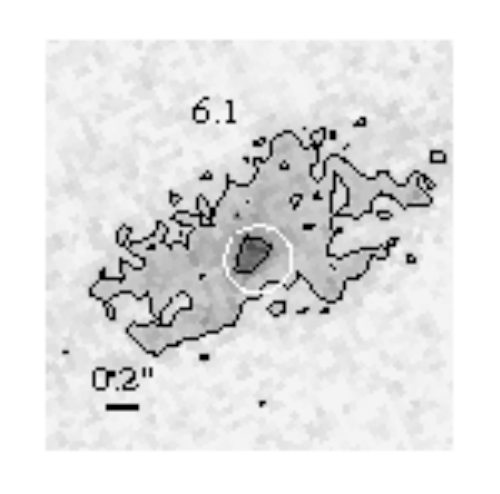}&
\includegraphics[width=0.20\linewidth]{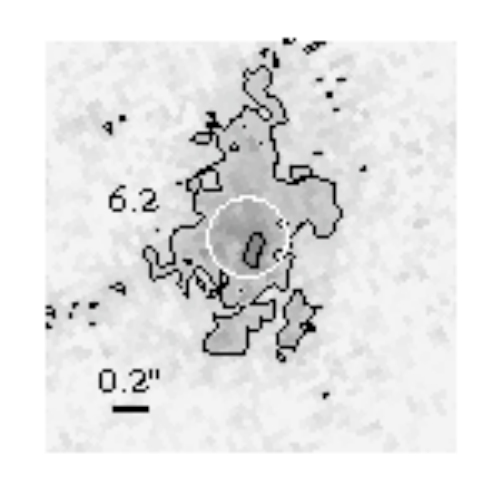}&
\includegraphics[width=0.20\linewidth]{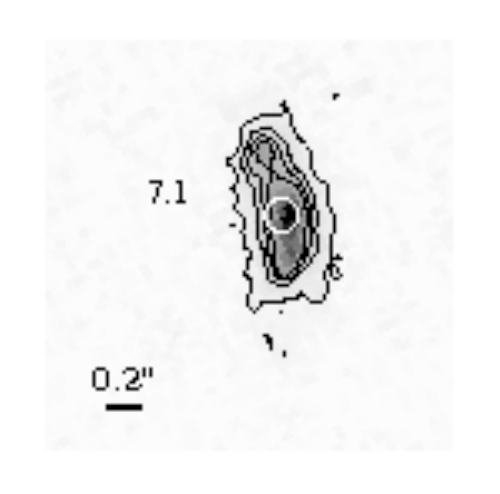}&
\includegraphics[width=0.20\linewidth]{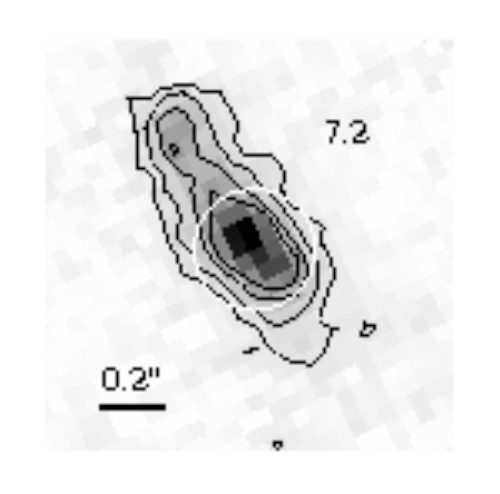}&
\includegraphics[width=0.20\linewidth]{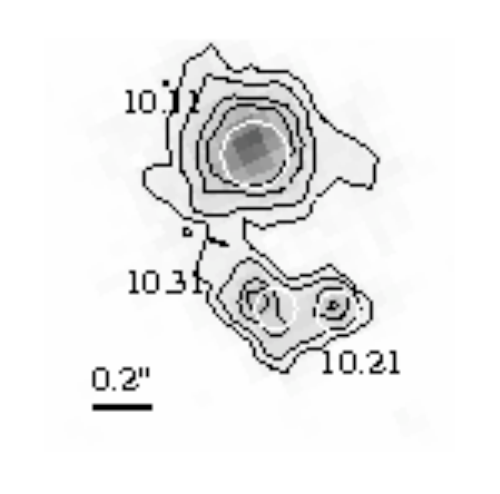}\\
\includegraphics[width=0.20\linewidth]{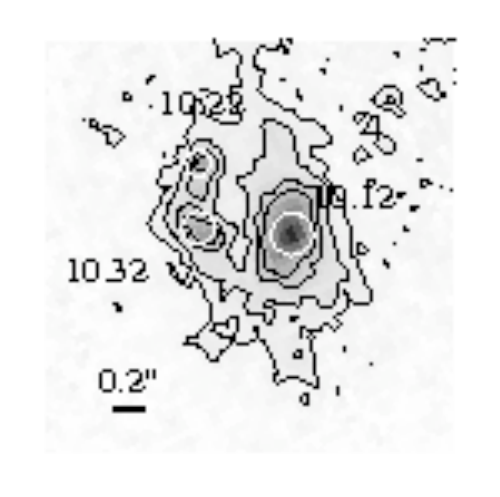}&
\includegraphics[width=0.20\linewidth]{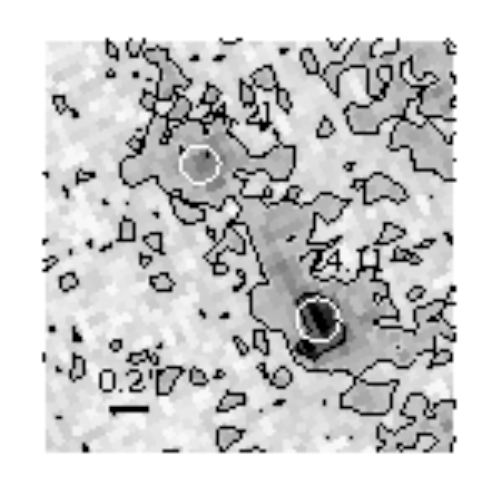}&
\includegraphics[width=0.20\linewidth]{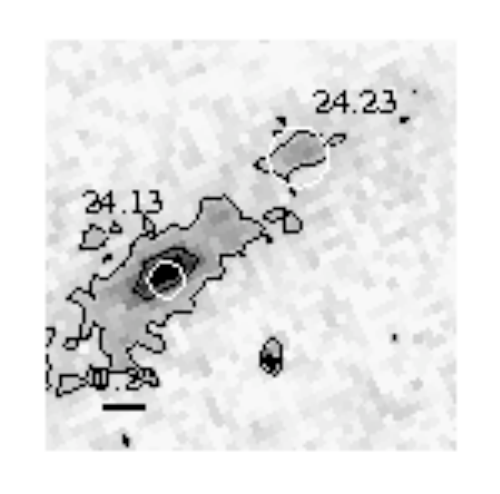}&
\includegraphics[width=0.20\linewidth]{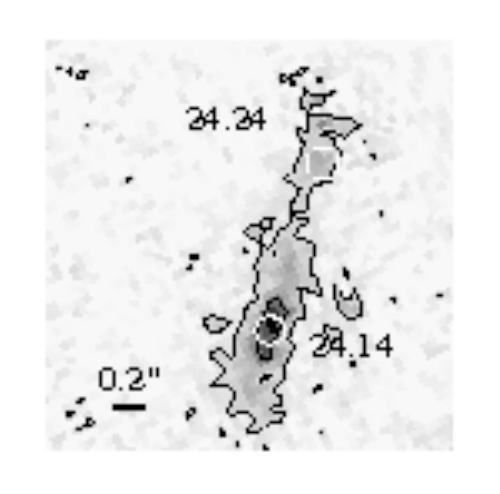}&\\
\end{tabular}
\caption{The identification of multiple images in Abell~1689 from the
high resolution {\it HST ACS} data: shown are a few examples wherein
the white circle plotted is the positional accuracy of each image. The
images are labeled as A.B where A denotes the family number and B the
image number within family A. Note that even within a given family of
images of the same source, the error varies due to the variation in S/N
of the images determined by the background that in turn depends on the
location of the images.}
\label{fig:image_fams}
\end{figure}

\begin{table}
\begin{tabular}{ccccccc}
ID    &  R.A. & Decl. & z & $\sigma_x$ & $\sigma_y$ & RMS \\
1.1   & 13:11:26.45 & -1:19:56.38 & 3.050  & 0.11 & 0.11 & 0.57 \\
1.3   & 13:11:29.77 & -1:21:07.32 & 3.050  & 0.10 & 0.10 & 0.17 \\
2.1   & 13:11:26.52 & -1:19:55.08 & 2.533  & 0.07 & 0.07 & 0.30 \\
2.4   & 13:11:29.81 & -1:21:05.96 & 2.533  & 0.11 & 0.11 & 0.34 \\
4.1a  & 13:11:32.17 & -1:20:57.34 & 1.1648 & 0.10 & 0.10 & 0.47 \\
4.1b  & 13:11:32.11 & -1:20:58.31 & 1.1648 & 0.07 & 0.07 & 0.44 \\
4.2a  & 13:11:30.52 & -1:21:11.90 & 1.1648 & 0.52 & 0.52 & 1.27 \\
4.2b  & 13:11:30.65 & -1:21:11.41 & 1.1648 & 0.05 & 0.05 & 0.48 \\
4.3a  & 13:11:30.76 & -1:20:08.01 & 1.1648 & 0.11 & 0.11 & 0.41 \\
4.4a  & 13:11:26.28 & -1:20:35.06 & 1.1648 & 0.11 & 0.11 & 0.27 \\
5.1   & 13:11:29.06 & -1:20:48.41 & 2.636  & 0.14 & 0.14 & 1.59 \\
5.3   & 13:11:34.11 & -1:20:20.88 & 2.636  & 0.10 & 0.10 & 0.27 \\
6.1   & 13:11:30.76 & -1:19:37.90 & 1.15   & 0.21 & 0.21 & 3.28 \\
6.2   & 13:11:33.35 & -1:20:11.97 & 1.15   & 0.24 & 0.24 & 1.23 \\
7.1   & 13:11:25.44 & -1:20:51.53 & 4.86   & 0.10 & 0.10 & 0.56 \\
7.2   & 13:11:30.67 & -1:20:13.80 & 4.86   & 0.19 & 0.19 & 0.46 \\
10.1a & 13:11:33.97 & -1:20:51.00 & 1.830  & 0.12 & 0.12 & 0.13 \\
10.1b & 13:11:33.95 & -1:20:51.55 & 1.830  & 0.07 & 0.07 & 0.21 \\
10.1c & 13:11:33.96 & -1:20:51.54 & 1.830  & 0.07 & 0.07 & 0.18 \\
10.2a & 13:11:28.05 & -1:20:12.29 & 1.830  & 0.13 & 0.13 & 0.22 \\
10.2b & 13:11:28.09 & -1:20:11.85 & 1.830  & 0.08 & 0.08 & 0.32 \\
10.2c & 13:11:28.09 & -1:20:12.24 & 1.830  & 0.10 & 0.10 & 0.27 \\
24.1a & 13:11:29.19 & -1:20:56.05 & 2.63   & 0.10 & 0.10 & 0.42 \\
24.1b & 13:11:29.22 & -1:20:55.28 & 2.63   & 0.09 & 0.09 & 0.31 \\
24.3a & 13:11:30.29 & -1:19:33.87 & 2.63   & 0.09 & 0.09 & 0.85 \\
24.3b & 13:11:30.25 & -1:19:33.27 & 2.63   & 0.15 & 0.15 & 1.31 \\
24.4a & 13:11:33.72 & -1:20:19.82 & 2.63   & 0.09 & 0.09 & 0.17 \\
24.4b & 13:11:33.69 & -1:20:18.80 & 2.63   & 0.10 & 0.10 & 0.11 \\
    \end{tabular}
    \caption{Catalog of 28 images used in this work to constrain
    cosmology. The columns are ID, R.A., Decl., z, x (positional
    uncertainty in CCD row), y (positional uncertainty in CCD column),
    RMS (error between predicted and observed image positions). All
    images have secure spectroscopic redshifts.}
\end{table}

\section*{Strong lensing modeling}

Although numerical simulations suggest that dark matter halos are well
described by the spherically averaged universal Navarro-Frenk-White
(NFW) density profile, in lensing studies we routinely prefer the
truncated Pseudo Isothermal Elliptical Mass Distribution (PIEMD)
potentials \citet{kassiola1993} for the extra degree of freedom in
allowing ellipticity and due to fact that these models have finite
mass. The 3-dimensional PIEMD density distribution is given by:
\begin{eqnarray}
\rho(r) = \frac{\rho_0}{(1+r^2/r_{core}^2)(1+r^2/r^2_{cut})}\,.
\end{eqnarray}
The PIEMD potential is parameterised by a central density, $\rho_0$,
related to the central velocity dispersion of the potential,
$\sigma_0$, and two characteristic radii that define the scales on
which the slope of the density profile changes. The profile is flat in
the inner region, then isothermal ($\rho \simeq r^{-2}$) between
$r_{core}$ and $r_{cut}$, and steeply decreases ($\rho \simeq r^{-4}$)
beyond $r_{cut}$. In addition, the ellipticity $\epsilon$ and the
position angle $\theta$ further characterise the shape of the
potential. The PIEMD potential has been extensively and successfully
used to model galaxy clusters
\citet{kneib1996,smith2005,limousin2007b}.

In the inner regions of clusters that are pertinent to the calculation
of strong lensing effects, cluster galaxies are mostly early-type
galaxies, and have been found to follow the scaling relations like the
Faber-Jackson relation
\citet{faber1976,wuyts2004,fritz2005}. Utilizing these observed
empirical scalings in our models, we assume that the masses of cluster galaxies 
scale with their luminosity $L$, via the following scaling relations:
\begin{equation}
r_{core} = r_{core}^\star (L / L^\star)^{1/2} \;, \quad 
r_{cut} = r_{cut}^\star (L / L^\star)^{1/2} \;, \quad
\sigma_0 = \sigma_0^\star (L / L^\star)^{1/4}
\end{equation}
\noindent where $L^\star$ is the luminosity of a typical galaxy in the
cluster; $r_{core}$ is fixed to an arbitrary small value of $0.15$
kpc;$r_{cut}^\star$ and $\sigma_0^\star$ depend on the properties of
the cluster and are left as free parameters. The small value of
$r_{\rm core}$ is in agreement with the results of
\citet{koopmans2006}, from the modeling of individual lensing galaxies
wherein it appears that early-type galaxies are isothermal in their
inner regions. According to these definitions, the halo mass is
proportional to $\sigma_0^2 r_{cut}$ and the mass-to-light ratio is
constant \citep{natarajan1997}.

Cosmological angular diameter distances are modeled according to the
definition from \citet{turner1997} as,
\begin{equation}
D_A = \frac{c H_0^{-1}}{1+z} \int_{z_1}^{z_2}\,dz  \left(
\Omega_{\rm m}(1+z)^3 + (1+z)^{3(w_{\rm x} + 1)} \Omega_{\rm x} \right)^{ 
-1/2}\,.
\end{equation}
In this equation, we assume a flat Universe (choice of prior). In
addition, since we assume matter and dark energy to be the only
significant components in the Universe, the sum of their respective
parameters $\Omega_{\rm m} + \Omega_{\rm x} = 1$. Finally, we assume
the Hubble constant to be $H_0\,=\,74\,{\rm km/s/Mpc}$
\citep{suyu2010}.  As for the two cosmological parameters
$(\Omega_{\rm m},w_{\rm x})$, we assign the following uniform priors
$\Omega_m \in [0,1]$ and $w_{\rm x} \in [-2,0]$.

In optical images, galaxies in Abell~1689 form 2 groups~: one
predominantly in the centre, and the second located about one
arcminute to the North-East. As in \citet{limousin2007b}, our final
model for Abell~1689 consists of 2 cluster-scale halos, one
galaxy-scale halo to model the BCG at the center of the dominant
group. On top of that, we add the color-magnitude selected catalog of
cluster member galaxies from \citet{limousin2007b}, but with an
additional cut in deflection angle.  Only halos that produce a
deflection larger than 0.07 arcsecond (i.e. about half the
observational uncertainty on the image positions) are selected. Thus,
we obtain a catalog of 58 cluster member galaxies brighter than $m_K <
18.11$. Each galaxy is assigned a PIEMD potential and the scaling
relations outlined above are adopted. In total, our final mass model
has a total of 21 free parameters, with 124 image constraints. The
adopted best estimate values for the mass model parameters are
reported in Table~2.

\begin{table}
\centering
\begin{tabular}{ccccccccc}
\hline
\multicolumn{1}{l}{Cluster} & $\Delta$\ R.A.$^a$ & $\Delta$\ Decl.$^a$ & $\epsilon^b$ &$\theta^c$ & $r_{core}$ & $r_{cut}$ & $\sigma_0$ & RMS \\ 
Mass component & (arcsec) & (arcsec) && (deg) & (arcsec) & (arcsec) & (km s$^{-1}$) & (arcsec) \\
\hline
\multicolumn{1}{l}{ Abell~1689~:} \\

Cluster \#1 &  \vp{0.5}{0.6}{0.5} & \vp{-4.2}{0.5}{0.6}  & \vp{0.32}{0.03}{0.02} 
& \vp{88.5}{1.6}{1.9} & \vp{14}{2}{1} & \sgl{1.5\ \mathrm{Mpc}} & \vp{1249}{29}{30}  & 2.54 \\

Cluster \#2 & \vp{-58.7}{1.5}{0.5} & \vp{43.2}{1.0}{3.5} & \vp{0.84}{0.00}{0.01} & \vp{125}{2}{3} 
& \vp{0.6}{0.0}{0.3} & \sgl{1.5\ \mathrm{Mpc}} & \vp{555}{26}{28} & {} \\

BCG & \vp{-6.1}{0.9}{0.0} & \vp{-4.5}{1.8}{0.0} & \sgl{0.13} & \sgl{85} & \vp{0.5}{0.0}{0.5} 
& \vp{14}{41}{0} &\vp{387}{71}{112} & {}\\

\multicolumn{1}{c}{\begin{minipage}{2cm}\centering L$^\star$ ral. \par\hspace*{\fill} $\to m_K^\star = 16$ \end{minipage} } &   \dots & \dots & \dots & \dots & \sgl{0.03}& \vp{26}{52}{15} & \vp{30}{188}{13} 
& {}\\ 
\hline
\end{tabular}
\caption{Estimated values and errors at 1$\sigma$ confidence for the
parameters of the mass model of Abell~1689; $^a$ the position of each
mass component is given relative to the cluster centre; $^b$ the
ellipticity corresponds to the {\it lenstool} input ellipticity for a
PIEMD potential $\epsilon = (a^2 - b^2)/(a^ + b^2)$, where $a$ and $b$ are
major and minor axis of the PIEMD mass distribution; the orientation 
$\theta$ increases from West to North.}
\end{table}

Before launching any computationally intensive runs with the full set
of images, we perform a quick check with a scaled down version of our
mass model excluding cluster member galaxies. For the two smooth
cluster-scale potentials and the BCG potential, we optimise the
positions, ellipticities, orientations, core radii, cut radii, and
velocity dispersions. We also let $\Omega_{\rm m}$ and $w_{\rm x}$
free to vary between the limits stated above.  As constraints, we use
two systems at redshifts $3.050$ and $2.533$. These are both 5-fold
multiply imaged systems with most of their images located far from any
bright cluster galaxy. In addition, these 2 systems form two Einstein
crosses, with most of their images appearing at the same distance from
the cluster center. Using these two systems, the estimated cluster
mass obtained is rather insensitive to our choice of the form of the
density profile. Consequently, our estimated values for $\Omega_{\rm
m}$ and $w_{\rm x}$ from these systems will be independent of our
detailed modeling of the lens potential. We obtain a degeneracy
between $\Omega_{\rm m}$ and $w_{\rm x}$ in agreement with our
theoretical expectations \citep{gilmore2009}. Although we obtain large
error bars on both $\Omega_{\rm m}$ and $w_{\rm x}$, the favored
cosmological parameter values ($\Omega_{\rm m} = 0.3$ and $w_{\rm x} =
-1$) are well centred in the 1$\sigma$ contour. For all subsequent
runs, we use the 3$\sigma$ confidence intervals of each mass model
parameter derived from this run as priors.

Note that the choice of parameterization of the mass model (NFW profile
versus PIEMD) does not impact recovery of cosmological parameters. While 
NFW profiles are found to be a good fit to simulated
massive galaxy clusters, in reality the NFW is not always the best fit
to lensing data of observed galaxy clusters. From our extensive
experience is modeling the mass distributions of lensing clusters
combining constraints from both the strong and weak lensing regime, we
find that for a majority of clusters the PIEMD (Pseudo-Isothermal
elliptical mass distribution) is a significantly better fit that the
NFW. This is most definitely the case for Abell 1689. In our earlier
work on modeling the mass distribution of this cluster presented in
Limousin et al. (2008), we investigated a range of models NFW,
generalized elliptical NFW, PIEMD and power law models. Ranking model
fits using Bayesian evidence derived by using the MCMC sampler we
found that the PIEMD model offered the best fit to the current
data. Therefore, as a starting point for the cosmography work we
started with the best-fit mass model parameters which are of course
then simultaneously optimised within reasonable ranges (the model is
not fixed).

In order to test the role of this choice of parametric model on the
recovery of cosmological parameters, we have performed detailed
simulations with many realizations of both the NFW and PIEMD modeled
clusters. As we show in Figure~3 the choice of mass model does not
bias recovery of input parameters. In the simulations, for both mass
distributions, it was assumed that we have constraints from 20 image
families all with measured spectroscopic redshifts.  The simulated
cluster are taken to be at the same redshift z = 0.2, each with 20
families created from the same background source catalog: one is PIEMD
and one is NFW.  There are no subclumps included as the issue to
settle here is the dependence of the choice of the smooth
model. Typical observational errors were assumed for both cases. It
turns out that the PIEMD contours are slightly wider, probably due to
the additional parameter needed.

\begin{figure}
\begin{center}
\resizebox{7.0cm}{!}{\includegraphics{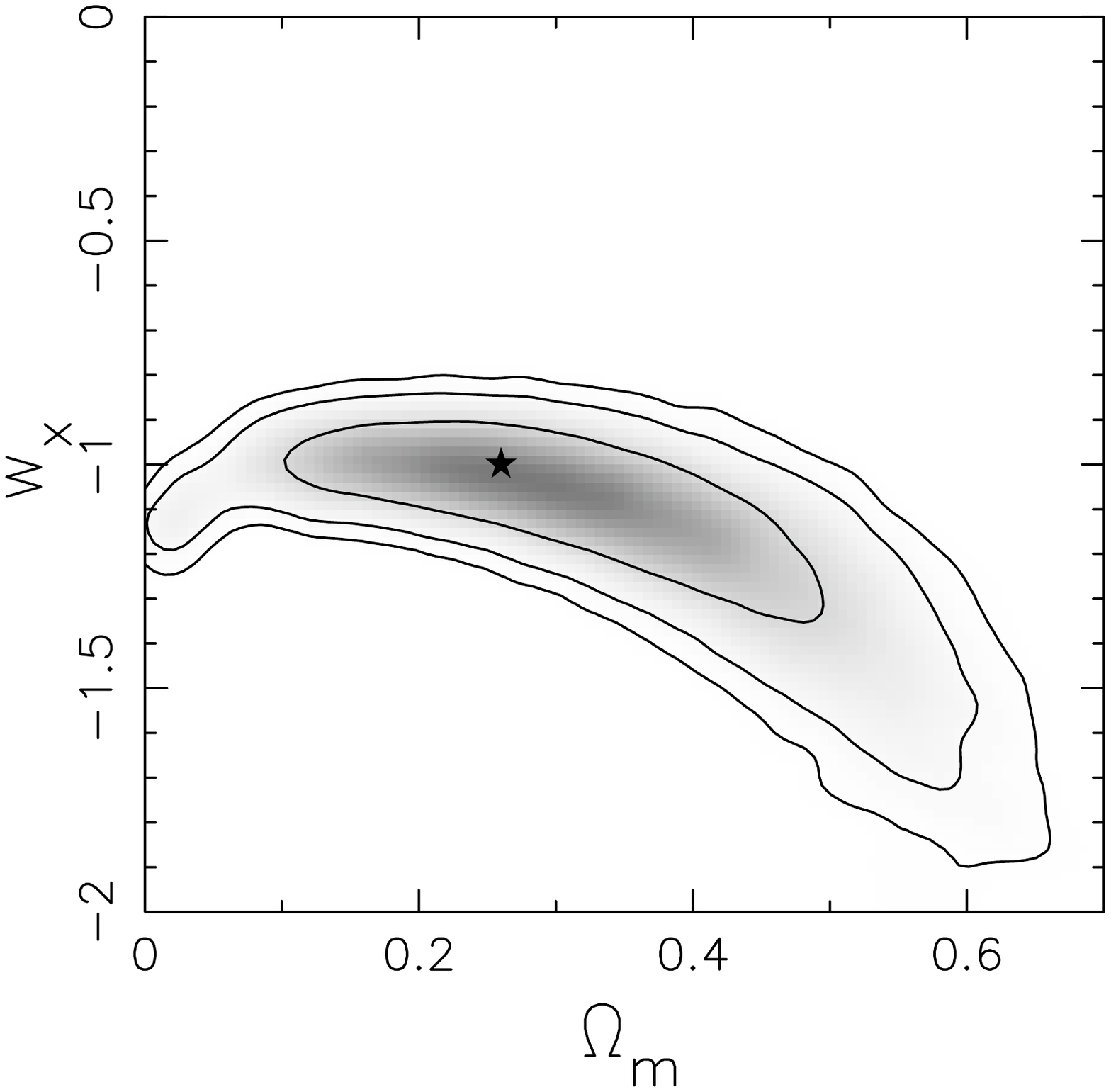}}\hspace{01.3cm}
\resizebox{7.0cm}{!}{\includegraphics{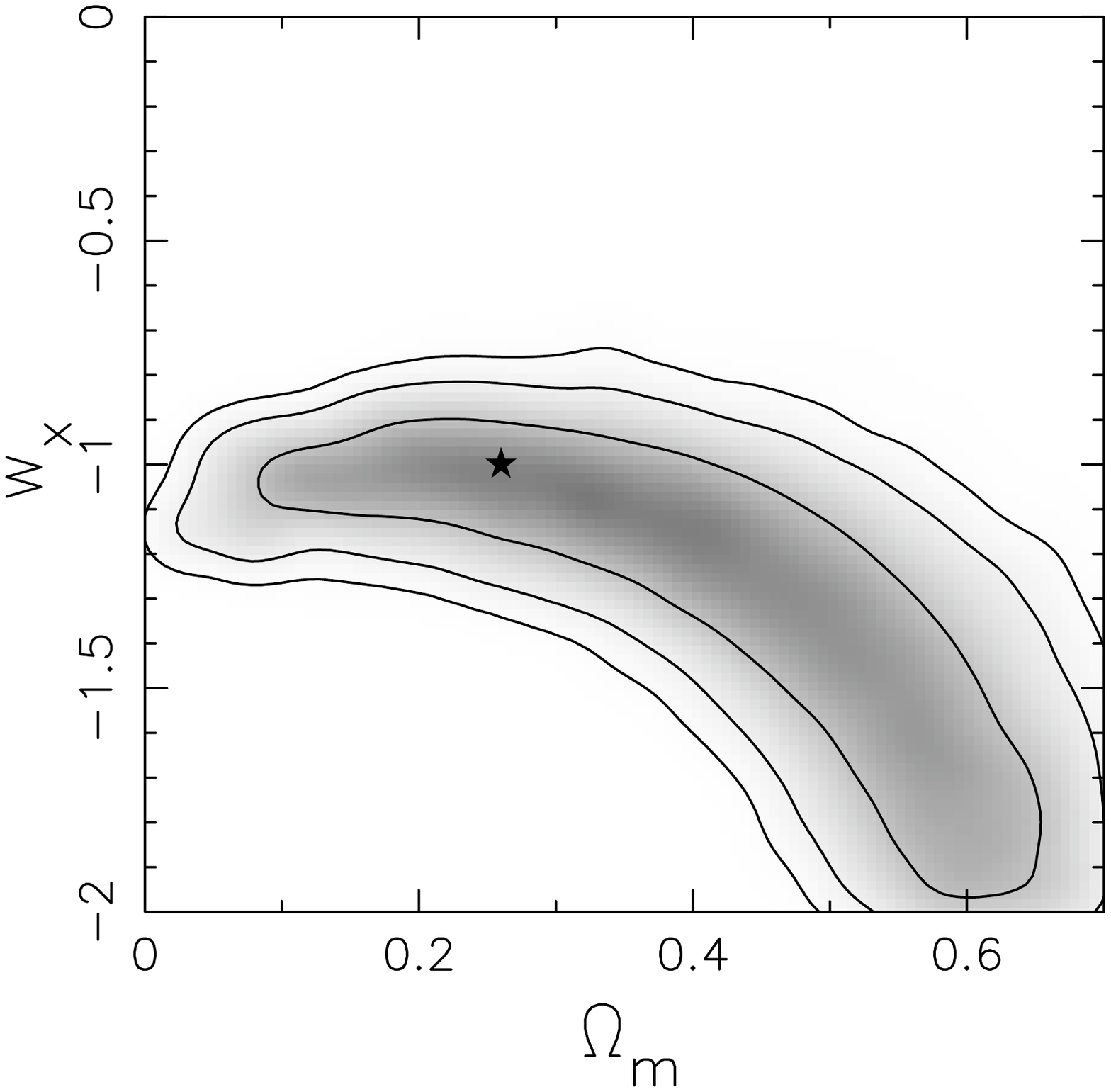}}
\end{center}
\caption{Left Panel: The recovery of input cosmological parameters 
with a simulated cluster modeled with an NFW profile utilizing 20 
image families. Right Panel: The recovery with the same cluster
and background image catalog as above but with a cluster modeled
with a PIEMD profile.}
\end{figure}

\begin{figure}
\includegraphics[width=0.99\linewidth]{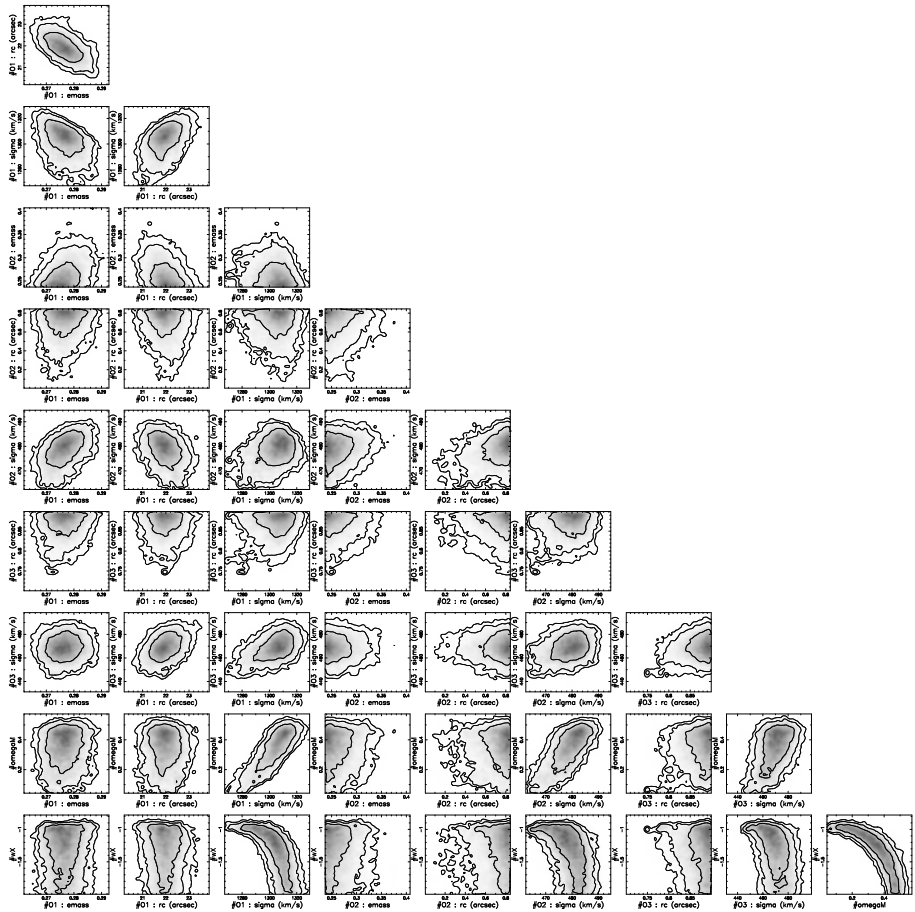}
\caption{Tabulation of the key parameter degeneracies: cosmological
parameters are degenerate with the following lens model parameters -
velocity dispersion, ellipticity and core radius. This is shown in the
results from the combined Bayesian optimization of the mass
distribution and cosmology in Abell~1689. In this plot, emass is the
ellipticity of the mass; $r_{\rm core} = r_{\rm c}$ the core radius
and sigma the velocity dispersion. The 2 large-scale clumps and the
BCG are denoted by $1$, $2$ and $3$ respectively.}
\label{fig:degen}
\end{figure}

\section*{Inventory of systematic errors in the modeling of Abell~1689}

Due to the large number of images found in Abell~1689, we can expect
the constraints on cosmology to be very tight. However, it is known
that parametric models are not flexible enough to reproduce all the
images at the observational level of accuracy
\citet{tomsubaru,oguri2005}.  Moreover, an initial analysis with the
0.13 arcsecond positional uncertainty found above, yielded different
modes in our parameter space with similar likelihoods that provided
very different estimates for $\Omega_{\rm m}$ and $w_{\rm x}$. This
often happens when systematic errors are not adequately taken into
account.  Therefore, we present a preliminary analysis of the
systematic errors for Abell~1689 in this work.  We treat two sources
in this work: systematic errors arising from cluster galaxies in the
lens plane and halos along the line of sight.

\flushleft{\bf Substructure in the lens plane:} The velocity
dispersions and scale radii of cluster member galaxies are likely to
display significant scatter about the scaling relations we assumed to
model them. If not properly accounted for, this scatter can introduce
biases into the parameter recovery.  In order to quantify this effect
on an individual image basis, we perform Monte Carlo simulations of
the lens system. We first obtain a set of source locations by mapping
each observed image back to the source plane using the initial cluster
model.  We randomly draw $\sigma_0$ and $r_{\mathrm{cut}}$ parameters
for the cluster galaxies from Gaussian distributions with mean values
obtained from the above scaling relations and standard deviations
equal to $20$ per cent of the mean.  We then lens the source locations
to the image plane using the cluster model with each realization of
the galaxy population.  We find that a $~20$ per cent scatter in the
velocity dispersions and scale radii can induce modeling errors on the
image positions that can be as large as $\sim1$ arcsecond. Put in
another way, in trying to avoid potential biases in the cosmological
parameters, we should not expect our simplified models to reproduce
image positions to within $\sim1$ arcsecond given the limitations on
modeling the galaxy population.

\begin{figure}
\begin{center}
\includegraphics{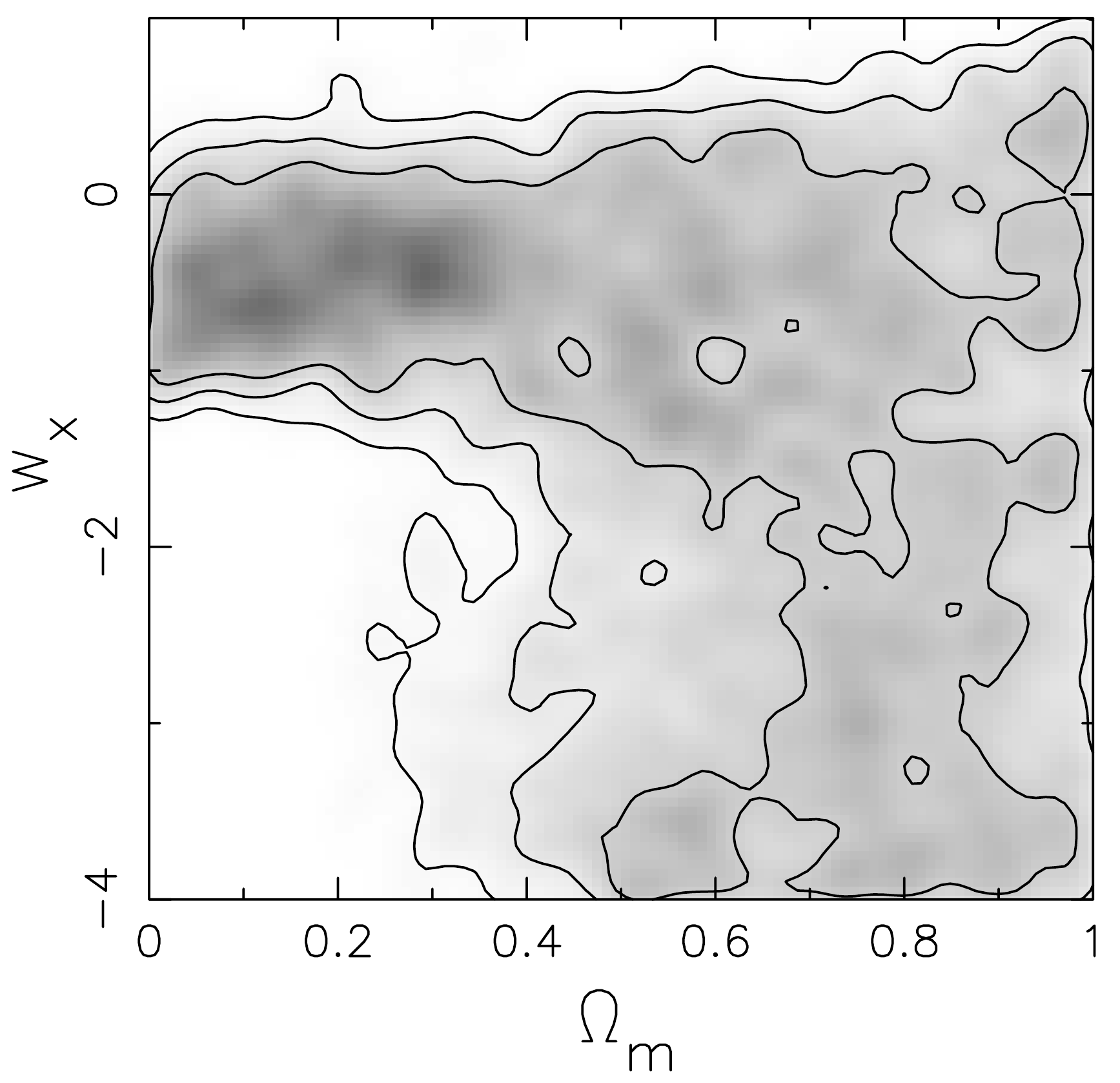}
\end{center}
\caption{The results from the simultaneous Bayesian optimization of
the detailed mass distribution and cosmological parameters in the
$\Omega_{\rm m}-w_{\rm x}$ plane for Abell~1689 including all
available families of multiple images as constraints including all
sources of error from observations, the line of sight structure as
well as errors in scaling relations employed to model cluster
galaxies. For this iteration all the 102 images were used including 58
cluster galaxies, error from all systematics were added to each
image. The RMS error in reconstructing image positions was large
signaling that the mass model was not accurate enough.}
\label{fig:combALL}
\end{figure}

\flushleft{\bf Substructure along the line of sight:} We also quantify
the errors due to halos along the line of sight using the Millennium
Simulation (MS) halo catalogs \citep{millenium}, which are freely
available for download.  We construct lens planes by selecting
randomly oriented slices from each MS snapshot.  We obtain a catalog
of halo positions and masses (down to $10^{11}\,M_{\odot}$) in each slice
and project their positions along one direction.  We then place
analytic NFW potentials at each halo location, using the mass and
redshift to assign a concentration parameter through scaling relations
obtained from N-body simulations in the literature.  The lens planes
are tiled along the line of sight and the cluster model is inserted at
the appropriate redshift. We ray trace the sources through many
line-of-sight realizations. We find that, although the line-of-sight
structure only occasionally changes the multiplicity of a family, the
more common effect is to simply perturb the image positions.  We find
that these perturbations are also typically on the order of 1
arcsecond.

We also study the effect of line of sight structure correlated with
the cluster on the recovery of cosmological parameters using
simulations. In our analysis, we utilize the multiple lensplane
approximation.  The total deflection is approximated as a series of
point deflections occurring at discrete redshifts.  The separation
between lensplanes is typically $\sim$100 Mpc.  In other words, all matter
within a rectangle of length $\sim$100 Mpc is projected onto a plane.  This
is generally known to be a satisfactory approximation to treating the
matter distribution as continuous.  Regarding the effect of correlated
large-scale structure, we note it can be reasonably modeled as
sitting at the redshift of the cluster.  Hence, it is reasonable to
treat any correlated structure within $\sim$50 Mpc as part of the projected
mass of the lens itself.  As an important example, we explored the
effect of a long filament aligned along the line-of-sight behind the
cluster. At the present, we are not able to utilize N-body
simulations for this exploration.  Instead, we use a simple parametric
approach.  We place a primary lens, representing a massive cluster (with
mass $M = 1.7 10^15\,M_{\odot}$), at $z = 0.184$.  We then place 5 equivalent circular
potentials, with centers at the origin, spaced 10 Mpc apart behind the
cluster to represent a filament (this is equivalent to breaking up the
filament into 5 lensplanes).  We use PIEMD profiles for these
potentials with core radii of of 0.5 Mpc and cut radii of 1.5 Mpc.
With these parameters, each contributing potential has a constant
density inside 0.5 Mpc and drops off as $r^{-2}$ for $r < 1.5$ Mpc.  Outside
of 1.5 Mpc, the density goes as $r^{-4}$.  The total mass within the
simulated filament is $10^{15}\,M_{\odot}$.  We also create a "projected" model
in which all potentials are placed on the same lensplane at $z = 0.184$.

We lensed 10 simulated source catalogs with both the projected and
full models and compared the resulting images. On average, deviations
between the full and projected models are around 0.6 arcsecond; less
than half of the typical combined error from cluster galaxy scatter
and uncorrelated LOS halos. So even in the worst-case scenario of a
long filament with mass roughly equal to the cluster, with its axis
aligned exactly along the line-of-sight, the error due to modeling it
in the same lensplane as the cluster is subdominant to the effects we
have considered in our analysis.

\section*{Bayesian MCMC approach to exploiting strong lensing observations}

We start with a parametrization describing the lens, the starting
point as ddescribed earlier is the best-fit model for Abell 1689
published by Limousin et al. (2008). Using LENSTOOL software we
explore the parameter space around the best-fit model while
simultaneously solving for the family ratio for pairs of images. This
is done while eproducing the location of the observed multiple images
within the supplied uncertainties. Earlier versions of the software
(Kneib et al. 1993) were based on a downhill minimization. However,
that technique was very sensitive to local minima in the likelihood
distribution; as a result, the modeling of complex systems as seen in
Abell 1689 became rapidly too inefficient. In order to tackle the
current observational data for Abell 1689, we implemented a new
optimization method based on a Bayesian Markov Chain Monte Carlo
(MCMC) approach. The initial assumed model defines the prior PDF from
which after sampling the posterior PDF is constructed for a range of
acceptable model parameters. We use the Bayesian evidence to rank the
models that best reproduce systems of multiple images. The Bayesian
evidence is used to rank and characterize models when scanning
parameter space. More details on this procedure can be found in Jullo
et al. (2008).

In terms of the actual chains used, we run 10 interlinked Markov
chains at the same time to prevent any Markov chain from falling into
a local minimum.  The MCMC convergence to the posterior PDF is
performed with a variant of the "thermodynamic integration" technique
(Ruanaidh and Fitzgerald 1996) called selective annealing. Here
"selective" stands for the following process. At each step, 10 new
samples (one per Markov chain) are drawn randomly from the current
posterior PDF (which corresponds to the prior PDF at the
beginning). These samples are then weighted according to their
likelihood and selected with a variant of the Metropolis-Hasting
algorithm (Metropolis et al. 1953; Hastings 1970). Roughly, the
samples with the worst likelihood are deleted and the ones with the
best likelihood are duplicated so that we always have 10 Markov chains
running at the same time. These chains are run in parallel for
efficiency. This ensures that the choice of initial seed does not
impact the calculation, it generates more samples yielding smoother
contour plots for the cosmological parameters $\Omega_{\rm m}$ and
$w_{\rm x}$.

\bibliographystyle{Science}
\bibliography{all}

\end{document}